\documentclass[]{emulateapj}

\usepackage{subfigmat}

\newcommand{\chisq}{$\chi^{2}$}

\newcommand{\figname}{Figure~}
\newcommand{\tabname}{Table~}

\newcommand{\age}{{\it age}}
\newcommand{\Z}{$Z$}
\newcommand{\efold}{$\tau_{e}$}
\newcommand{\ebv}{$E(B-V)$}

\newcommand{\largesample}{$\ga 10^{5}$}
\newcommand{\ND}{$N$}
\newcommand{\numcorners}{$2^{N}$}
\newcommand{\numneighbors}{\numcorners}
\newcommand{\Ngrid}{$N_{\mathrm{grid}}$}

\newcommand{\sn}{$S/N$}

\newcommand{\nrealization}{20}
\newcommand{\restwaverange}{$3450-8350$~\AA}
\newcommand{\rangeAge}{$1.0, 4.2, 7.3, 10.5, 13.7$}
\newcommand{\rangeMetallicity}{$0.0004, 0.004, 0.008, 0.02, 0.05$}
\newcommand{\rangeTau}{$1.0, 4.5, 8.0, 11.5, 15.0$}
\newcommand{\rangeEbv}{$0.0, 0.2, 0.4, 0.6, 0.8$}
\newcommand{\targetparm}{8~Gyr, 0.02, 9~Gyr, 0.3~mag}
\newcommand{\ninitialmodel}{625}
\newcommand{\nsynthetic}{100}
\newcommand{\snsynthetic}{20}
\newcommand{\unbiasednspec}{$\ga 10^4$}
\newcommand{\examplerchisq}{$1.07$}
\newcommand{\avgrchisq}{$1.0$}
\newcommand{\avgferr}{$3.5\%$}
\newcommand{\minferr}{$0.4\%$}
\newcommand{\maxferr}{$11.3\%$}
\newcommand{\nspecC}{320}
\newcommand{\rangeAgeC}{$3.0, 6.0, 8.0, 12.0$}
\newcommand{\rangeMetallicityC}{$0.0004, 0.004, 0.008, 0.02, 0.05$}
\newcommand{\rangeTauC}{$3.0, 6.0, 9.0, 14.0$}
\newcommand{\rangeEbvC}{$0.1, 0.3, 0.5, 0.7$}

\begin{document}

\title{Generating on-the-fly large samples of theoretical spectra through \ND-dimensional grid}

\author{Ching-Wa Yip}

\altaffiltext{}{Department of  Physics and Astronomy,  The Johns Hopkins
  University, Baltimore, MD 21218, USA.}

\email{cwyip@pha.jhu.edu}

\begin{abstract}
Many  analyses  and  parameter  estimations  undertaken  in  astronomy
require  a  large set  (\largesample)  of non-analytical,  theoretical
spectra, each  of these defined  by multiple parameters.   We describe
the  construction of  an \ND-dimensional  grid which  is  suitable for
generating  such spectra.   The  theoretical spectra  are designed  to
correspond  to  a targeted  parameter  grid  but  otherwise to  random
positions in the parameter space, and they are interpolated on-the-fly
through  a  pre-calculated  grid  of  spectra.  The  initial  grid  is
designed to  be relatively  low in parameter  resolution and  small in
occupied hard disk space and therefore can be updated efficiently when
a  new model  is  desired.  In  a  pilot study  of stellar  population
synthesis  of  galaxies,  the  mean  square errors  on  the  estimated
parameters are  found to decrease  with the targeted  grid resolution.
This  scheme of generating  a large  model grid  is general  for other
areas of studies, particularly  if they are based on multi-dimensional
parameter space and are focused on contrasting model differences.
\end{abstract}


\section{Introduction}

Various analyses and parameter estimations performed in growing number
of applications  in astronomy require  a large set  of non-analytical,
theoretical spectra,  where each spectrum in-principle  can be defined
by multiple  free parameters.   In stellar population  synthesis using
the                          Bayesian                         approach
\citep[e.g.,][]{2003MNRAS.341...33K,2005MNRAS.362...41G,2007ApJS..173..267S},
as many as $10^{4} -  10^{5}$ theoretical spectra and the derived line
indices are  used in  order to  cover a wide  range of  star formation
histories from  early- to  late-type galaxies.  With  about 5  or more
free parameters,  each theoretical spectrum is a  single stellar burst
defined by metallicity and age,  superimposed at a given mass fraction
on a  spectral component with exponentially  decreasing star formation
rate characterized by  the age of the oldest  stars, e-folding time of
star formation, and both spectral components could be dust-attenuated.
Similarly,  $10^{5} -  10^{6}$ theoretical  spectra are  considered in
parameter estimation on stellar spectra \citep[e.g.,][who considered a
model   defined  by   14   parameters]{2007ApJS..169..328R},  and   on
\ion{H}{2}  regions \citep[e.g.,][who  studied a  model defined  by 15
parameters]{2009MmSAI..80..397M}.  One approach  to prepare and manage
a non-analytical model is to  store all of the pre-calculated spectra,
fixed at both  the parameter choice and the  parameter resolution.  As
models are  becoming more sophisticated and are  growing in varieties,
however, updating and storing $10^{5}$  or more spectra may not be the
most flexible approach  for many astronomers. To handle  large sets of
theoretical spectra at  multi-dimensional parameter space is therefore
a question that cannot be neglected.

We  consider here current  integrated stellar  population models  as a
case  study.    These  models,  together   with  parameter  estimation
techniques  or  other analyses,  were  shown  by  many authors  to  be
invaluable for deriving composition and star formation rate/history of
different types of galaxies through their observed spectra, regardless
whether     or    not     a     large    model     grid    is     used
\citep[e.g.,][]{2000AJ....119.1645T,2001MNRAS.327..849R,2003MNRAS.341...33K,2003MNRAS.343.1145P,2003ApJ...587...55G,2004MNRAS.355..273C,2004ApJ...613..898T,2004MNRAS.351.1151B,2005MNRAS.358..363C,2005MNRAS.362...41G,2006MNRAS.365..385M,2006MNRAS.365...74O,2007MNRAS.378.1550P,2007MNRAS.381..263A,2007MNRAS.381.1252T,2009MNRAS.393..406C,2009MNRAS.tmp.1327R}.
Firstly,  there  is  a  variety of  spectral  synthesis  computational
programs which use different flavors in initial stellar mass function,
stellar     types,      and/or     stellar     evolutionary     tracks
\citep[e.g.,][]{1997A&A...326..950F,1999ApJS..123....3L,2003MNRAS.344.1000B,2005MNRAS.362..799M,2007MNRAS.382..498C}. The
variations in input  ingredients exist not only among  the models, but
also  within a  single  model in  the  form of  additional freedom  in
parameter  choice.  Secondly, increasingly  more parameters  are being
included in  the analyses, meaning  that the parameter  space defining
the  theoretical   spectra  is  getting  larger.   So   much  so  that
\citet{2009ApJ...694..902L}   have  recently  constructed   a  stellar
population model  to the level of individual  element abundances, with
the goal  to understand the  effect of each  and every element  to the
integrated spectrum of a  stellar population.  However, there are very
few studies in the astronomy literature which address the case where a
multi-dimensional  parameter  space  defines  a  model.   Two  of  the
fundamental questions  are: (1)  how do we  obtain, or  interpolate, a
theoretical  spectrum   in-between  the  parameter   values,  that  is
originally  unavailable?   (2)  what  is  the best  way  for  such  an
interpolation, taking into  account of the noise in  the data, and the
possible non-linear dependence \citep{2009AJ....138.1365V} between the
theoretical spectrum  and its underlying physical  parameters?  One of
the goals of this work is to address the first question.

Here we describe  a novel approach which increases  the flexibility of
handling  a large set  (\largesample) of  theoretical spectra,  and is
general  for  \ND-dimension (\ND{D})  parameter  space.  The  approach
adopts a  hypercube, an \ND{D}  analog of a  cube of length  unity for
each side \citep[e.g.,][]{Anthony87}, to represent the parameter space
underlying each  and every theoretical spectrum.  The  spectrum at any
intermediate parameter point is generated on-the-fly\footnote{We refer
the on-the-fly ability to be  the generation of theoretical spectra at
run time  as required  by the actual  computation, e.g.,  in parameter
estimation.} through multi-linear interpolation, upon an initial model
grid  that  is  designed  to  be lower  in  parameter  resolution  and
therefore  can be  updated efficiently.   We examine  a pilot  case in
stellar population synthesis of galaxies, to show that the approach is
applicable to studies  in astronomy, and well manageable  by a typical
personal computer -- a computer with Intel(R) Pentium(R)D 3.20~GHz CPU
and 3.19~GHz RAM is used in this work.

All of  the spectra  considered in this  work are expressed  in vacuum
wavelength, and are re-sampled  to within the optical wavelength range
\restwaverange \ at  a resolution of 1~\AA \  per wavelength bin.  The
Cartesian  coordinate   system  is   used  throughout  this   work  in
representing the parameter space.

\section{Method: An \ND-dimensional grid for model parameters} 

Several  desirable characteristics  are  identified in  the setup  for
generating  a large model  grid: (1)  generality for  \ND{D} parameter
space; (2) the theoretical  spectra defined by the targeted parameters
are generated on-the-fly, which is essential for improving flexibility
and saving  disk space, especially in arbitrarily  complex models; (3)
targeted  parameters can  be specified  at arbitrary  position  in the
parameter  space  at  run  time,  with  the  goal  to  obtain  in  the
analysis/parameter  estimation an  arbitrary  computational resolution
which is limited only by the parameter resolution in the initial model
grid.

To fulfill the  above criteria, we generate a  small- to moderate-size
sample  of   theoretical  spectra  to  start   with.   The  underlying
parameters of this initial sample is  in the form of a Cartesian grid.
For each  parameter the size of the  bins is allowed to  be uneven, in
which case  the grid will be  rectangular instead of square  in the 2D
analogy.  For convenience we called this sample the {\it initial grid}
of theoretical spectra, the purpose of which is to form a leverage for
generating a larger sample of  spectra, the {\it targeted sample}, for
which the  defining parameters can be located  anywhere in-between the
initial grid points. If this sample is also decided to be described by
a parameter grid instead of  random points within the initial grid, we
called  that the  {\it targeted  grid} of  spectra.  The  advantage of
having  a targeted  grid  of  sample is  their  offering an  intuitive
setting for  performing necessary integrations over  a parameter space
in the  applied parameter estimation technique, an  approach also used
by                           various                           authors
\citep[e.g.,][]{2003AJ....126.2330C,2003MNRAS.341...33K,2005MNRAS.362...41G,2008ApJ...677..970W}.

The theoretical spectrum at  any intermediate grid point is calculated
by  first  locating the  corresponding  neighboring  spectra from  the
initial grid.   The neighboring parameter grid points  are then mapped
into the corners  of a hypercube of dimension \ND, where  \ND \ is the
number  of parameters defining  both the  initial and  targeted grids.
The   details   of   constructing   a  hypercube   is   described   in
\S\ref{section:hypercube}.   Multi-linear interpolation  (described in
\S\ref{section:MLinterp})  is  next   used  to  derive  the  concerned
theoretical value at  the targeted parameter point.  The  value is, in
the  current   context,  the  flux  density  in   a  given  wavelength
bin. Therefore,  the above  procedure is repeated  for each  and every
wavelength bin of interest. The whole procedure of model generation is
joined seamlessly with the actual computational routines for parameter
estimation or  other analyses, with  the only input being  the initial
model grid.

\subsection{Hypercube} \label{section:hypercube}

\begin{figure}
\includegraphics[width=88mm]{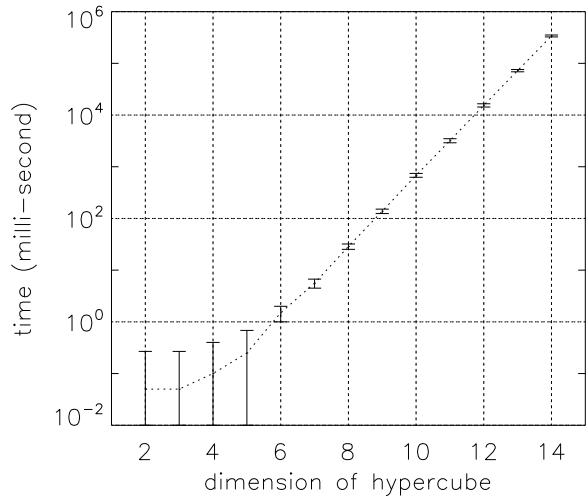}
\caption{The time for constructing a hypercube of dimension \ND, which
  is  used to represent  the number  of parameters  in the  model. For
  dust-attenuated   integrated   stellar   spectra   defined   by   an
  exponentially decreasing star  formation rate, \ND \ =  4 (\age, \Z,
  \efold,  \ebv),  the  time  is  less than  1~milli-second.   For  an
  additional dust-attenuated stellar burst, \ND  \ = 8 (\age, \Z, \ebv
  \  and the  mass  fraction of  the  burst), the  time  is less  than
  1~second.  Each error bar shows  the $\pm$ 1-sigma sample scatter of
  the required  time, after \nrealization \  realizations.  The dotted
  lines are for eye-guiding only.}
\label{fig:timehypercube}
\end{figure}

A hypercube is an \ND{D} cube with each side equals unity.  In 2 and 3
dimensions, a  hypercube is hence  a square and a  cube, respectively.
This  geometry  makes  a   hypercube  to  be  naturally  suitable  for
describing  an  \ND{D}  Cartesian  space,  such as  a  grid  of  model
parameters.  For our purpose, the  dimension of the parameter space is
defined  to  be  the  number  of parameters  that  fully  specifies  a
spectrum.   For example,  in a  simple stellar  population  defined by
stellar  age and  stellar  metallicity, \ND  \  = 2.   In practice,  a
hypercube is generated by specifying all the corners using vectors, or
arrays in  the actual computational  routines. For example, in  2D the
arrays are  $\{0, 0\}, \{0, 1\},  \{1, 0\}$ and $\{1,  1\}$, where any
coordinate value is defined  to be either 0 or 1.  It  is easy to show
by  deduction that  there are  in total  \numcorners \  corners  in an
\ND{D} hypercube.

\begin{figure}
\includegraphics[width=84mm]{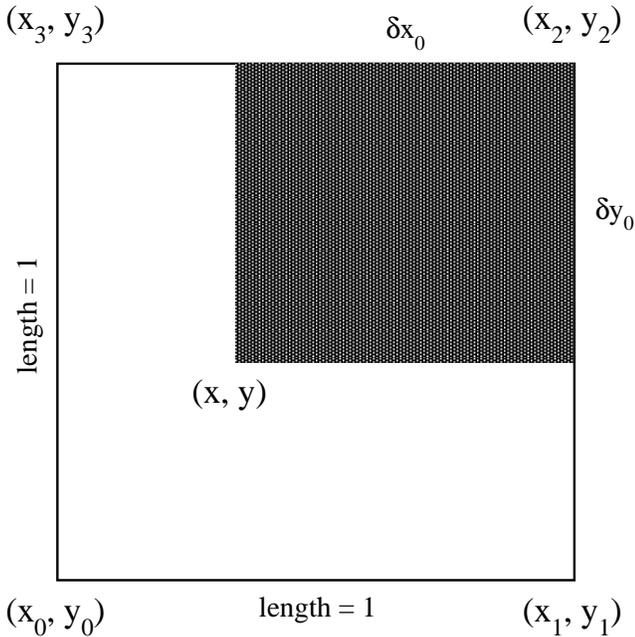}
\caption{An illustration  of the weighting scheme  in the multi-linear
interpolation, in a  2D hypercube (i.e., a square)  which encloses the
targeted point  $(x, y)$, under the Cartesian  coordinate system.  The
weight of  point $(x_{0}, y_{0})$ to  the targeted point  value is the
area of the stippled square, $\left| \delta{x}_{0} \cdot \delta{y}_{0}
\right|$.  When the  point $(x, y)$ is moved  toward $(x_{2}, y_{2})$,
the stippled area  shrinks, to the point that  when $(x, y)$ coincides
with $(x_{2}, y_{2})$, $\delta{x}_{0} \cdot \delta{y}_{0} = 0$, or the
weight on  the targeted point value  due to point  $(x_{0}, y_{0})$ is
zero.  Applying to theoretical simple stellar population, for example,
the $x$ direction can be \age, the $y$ direction be \Z.}
\label{fig:MLinterp}
\end{figure}

We  use a  random rendering  of points,  together with  a book-keeping
approach, to  create each  and every corner  of the  hypercube exactly
once.  After some algebraic  derivations in the low-dimensional cases,
we found that  the number of combinations -- or  the number of corners
-- in the  hypercube of dimension $N$,  for $k$ occurrences  of 1's in
the array specifying the corner is given by the binomial coefficient

\begin{equation}
{}_{N}C_{k} = \frac{N!}{k! \, (N-k)!} \ .
\end{equation}

\noindent
In 3D,  there is  1 ($=  {}_{3}C_{0}$) combination of  no 1:  $\{0, 0,
0\}$, 3 ($= {}_{3}C_{1}$) combinations  of one 1: $\{0, 0, 1\}$, $\{0,
1, 0\}$  and $\{1,  0, 0\}$, 3  ($= {}_{3}C_{2}$) combinations  of two
1's:  $\{0, 1,  1\}$,  $\{1, 1,  0\}$ and  $\{1,  0, 1\}$,  and 1  ($=
{}_{3}C_{3}$)  combination  of  three   1's:  $\{1,  1,  1\}$.   Other
approaches certainly  can be used  to accomplish the same  purpose, as
long  as   a  hypercube  is  automatically  created   given  only  its
dimensionality, \ND.

The time  required to  construct a hypercube  of a given  dimension is
shown   in   \figname\ref{fig:timehypercube}.    For   dust-attenuated
integrated  stellar  spectra defined  by  an exponentially  decreasing
star formation  rate, \ND  \ =  4 (\age,  \Z, \efold,  \ebv,  at fixed
initial stellar mass function and  other parameters), the time is less
than 1~milli-second.  For an extra dust-attenuated stellar burst which
is added at a  certain flux fraction, \ND \ = 8  (\age, \Z, \ebv \ and
the burst  mass relative to the  total galaxy mass), the  time is less
than 1~second.  The construction of the hypercube therefore should not
constitute as  bottleneck in typical analyses  or parameter estimation
problems, which take minutes to hours of computation.

\subsection{Multi-linear interpolation at intermediate grid point}
  \label{section:MLinterp}

To interpolate  a spectrum at  an intermediate parameter point  in the
initial grid, called the targeted  point, we first perform a search to
locate all of points, or the neighboring points, from the initial grid
that encloses  the targeted  point.  This search  is performed  on the
parameter-to-parameter basis, for each parameter a pair of neighboring
points  is  obtained.  As  a  result,  the  searching time  scales  as
\ND$\cdot$\Ngrid \ and is  quick in typical applications, where \Ngrid
\ is  the number of grid points  for a given parameter.   Next, we map
the corners of a  hypercube to these neighboring points, \numneighbors
\ in total for an \ND{D} hypercube.

The weight of the theoretical value  of a grid parameter point to that
in    the    targeted    parameter    point    is    illustrated    in
\figname\ref{fig:MLinterp},  in  the  2D  case. The  weight  of  point
$(x_{0}, y_{0})$  to the  targeted point is  the area of  the stippled
square,  $|\delta{x}_{0} \cdot \delta{y}_{0}|$.   When the  point $(x,
y)$   coincides   with    $(x_{2},   y_{2})$,   $\delta{x}_{0}   \cdot
\delta{y}_{0} = 0$,  or the weight on the targeted  point value due to
point $(x_{0}, y_{0})$  is zero.  The final interpolated  value at the
targeted point is  calculated by summing up contributions  from all of
the points

\begin{eqnarray}
\hat{f}(x, y) & = & {} f(x_{0}, y_{0}) \cdot \left| \delta{x}_{0} \cdot
\delta{y}_{0} \right| \nonumber \\
        & & + f(x_{1}, y_{1}) \cdot \left| \delta{x}_{1} \cdot
\delta{y}_{1} \right| \nonumber \\
        & & + f(x_{2}, y_{2}) \cdot \left| \delta{x}_{2} \cdot
\delta{y}_{2} \right| \nonumber \\
        &  &  +  f(x_{3},  y_{3}) \cdot  \left|  \delta{x}_{3}  \cdot
\delta{y}_{3} \right| \ .
\end{eqnarray}

\noindent
For  a  square  of  unit  length  for  each  side,  we  can  re-write
        $|\delta{x}_{0} \cdot \delta{y}_{0}|$ to be $(1 - |x - x_{0}|)
        \cdot (1 - |y - y_{0}|)$, similarly for other corners.

To extend to the \ND{D} parameter space, the interpolation formula for
the   value   of   interest   at   the   targeted   point   $\vec{x}$,
$\hat{f}(\vec{x})$, is

\begin{equation}
\hat{f}(\vec{x})  =  \sum_{i   =  1}^{2^N}  \,  f(\vec{z}{\,}^{i})  \,
w^{i}(\vec{x}) \ , \label{eqn:MLinterp}
\end{equation}

\noindent
where  $i$  denotes  the  sum  over  contributions  from  all  of  the
neighboring points,  $\vec{z}{\,}^{i}$, enclosing the  targeted point,
$\vec{x}$, and the weight of each neighboring point is

\begin{equation}
w^{i}(\vec{x})  =  \prod_{j  =  1}^{N}  \,  (1  -  {\left|  {x_{j}}  -
z^{i}_{j}\right|}) \ , \label{eqn:weight}
\end{equation}

\noindent 
in which  $j$ denotes a given  parameter axis in  the \ND{D} parameter
space.  For  a hypercube of length  unity for each  side, $\sum_{i} \,
w^{i}$  is equal to  unity.  This  formula is  an extension  of linear
interpolation in the 1D case by \citet[][his Eqn.~5.2]{Burnett86}, and
is  shown   to  be  applicable  to  computations   in  fluid  dynamics
\citep{2004MAN}.

The  Eqns.~\ref{eqn:MLinterp} and  \ref{eqn:weight}  assume normalized
coordinates, as  such the range  of each coordinate value  lies within
$[0,  1]$.    Since  the  neighboring   points  $\vec{z}{\,}^{i}$  are
represented by the corners of a hypercube, no extra normalization step
is required for those.  The target point is normalized according to

\begin{equation}
x_j = \frac{p_j - p_j^{-}}{p_{j}^{+} - p_{j}^{-}} \ , \label{eqn:norm}
\end{equation}

\noindent
where $p_{j}$ is the actual parameter  value (e.g., \age \ = 3~Gyr) at
point $x_{j}$, and $p_{j}^{-}$ and $p_{j}^{+}$ are that of the pair of
neighboring  points  enclosing  the  targeted point,  for  the  $j$-th
parameter axis (e.g.,  \age \ = 1~Gyr and  4.2~Gyr, respectively).  To
derive    the    flux    densities    of    a    targeted    spectrum,
Eqns.~\ref{eqn:MLinterp}  and  \ref{eqn:weight}  are  applied  on  the
wavelength-to-wavelength basis. If the uncertainty in the flux density
of the theoretical spectra is also available, Eqn.~\ref{eqn:MLinterp},
together  with  usual  error  propagation  formulae, can  be  used  to
obtained the uncertainty in the interpolated spectrum.

\begin{figure*}
\includegraphics[width=180mm]{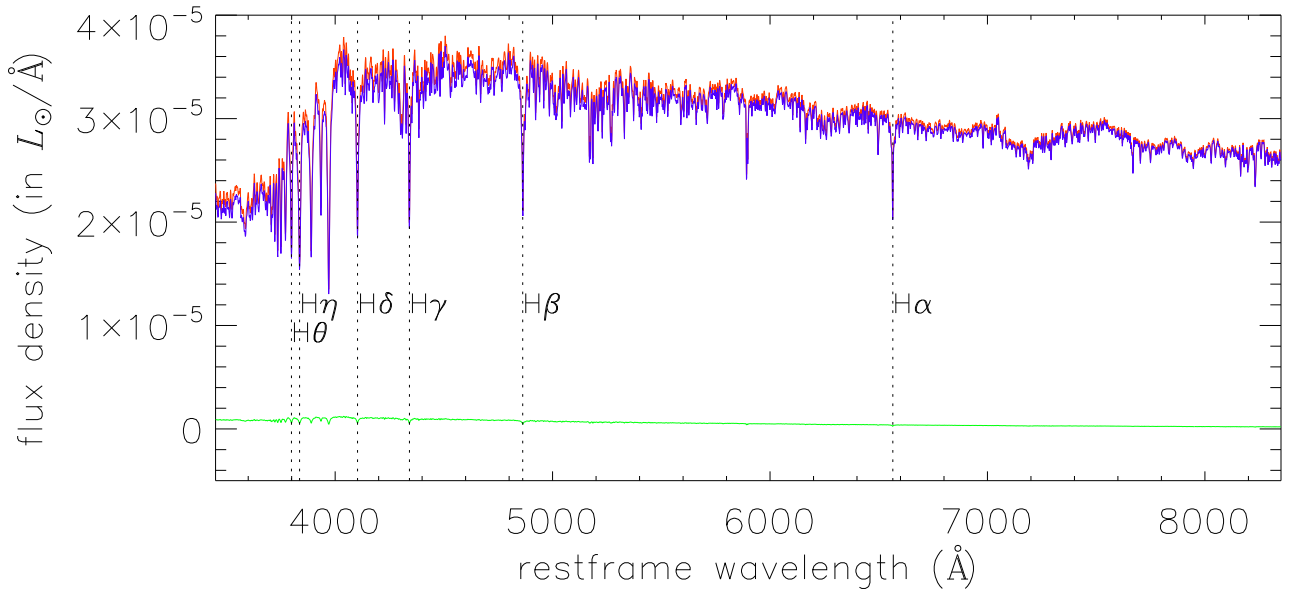}
\caption{An example interpolated spectrum (vacuum wavelength) by using
a 4D hypercube,  shown in red. The true  theoretical spectrum from the
dust-attenuated  \citet{2003MNRAS.344.1000B}   model  is  plotted  for
comparison,  in blue.   The difference  spectrum (interpolated  - true
theoretical)  is shown  in  green. Some  Balmer  absorption lines  are
marked.    The   solar  luminosity   $L_{\odot}$   is  $3.826   \times
10^{33}$~ergs~s$^{-1}$.   The flux density  error in  the interpolated
spectrum is  about 2\%, assigned to  give a reduced \chisq  \ close to
unity.}
\label{fig:demointerpspec}
\end{figure*}

The accuracy  of the interpolated  value at an  intermediate parameter
point  in-between the  initial grid  points  depends on  how good  the
linearity  assumption is.   This simplifying  assumption is  made only
locally,  i.e., in-between two  grid points  in the  initial parameter
grid, for  a given  parameter axis. There  is no requirement  that the
linearity has to stand in the global range of a parameter.  The reason
is that any non-linear change  of the theoretical value with parameter
globally  can  be taken  into  account,  e.g.,  by adopting  a  higher
sampling fraction of the initial  grid points in the related parameter
regions.  If  indeed the spectral features  depend highly non-linearly
with   parameters  that   happen   to  occur   at  unknown   parameter
amplitude(s), a preliminary step can  be taken to locate the concerned
parameter   amplitudes.   The   recently  introduced   locally  linear
embedding  \citep{2009AJ....138.1365V}   appear  to  be   a  promising
approach  for this  purpose, for  its ability  to unfold  a non-linear
manifold (see their \figname~1).

\section{Effect of Grid Resolution on Parameter Estimates} \label{S:gridres}

A fundamental  aspect in  parameter estimation is  the choice  of grid
resolution.   A  finer  parameter  grid  is expected  to  give  higher
accuracy  in  the  estimates,  whereas  a smaller  one  requires  less
computational time.  We  use here the on-the-fly ability  of our model
generation  approach to investigate  the effect  of the  targeted grid
resolution on  the parameter estimates.  This  study requires multiple
sets of models of increasingly large sample size.

To construct  an initial  grid we use  the \citet{2003MNRAS.344.1000B}
stellar population model (resolution 1~\AA \ per wavelength bin within
the    vacuum    wavelength    range    \restwaverange),    and    the
\citet{2000ApJ...533..682C} intrinsic dust model.  The initial stellar
mass function is that by \citet{2003PASP..115..763C} \citep[similar to
][with a turnover below $\sim0.3~M_{\odot}$]{1990MNRAS.244...76K} with
lower and upper limits in  stellar mass, $0.1 - 100~M_{\odot}$. A wide
range of stellar population and  dust parameters is covered -- the age
of the  oldest stars, \age,  takes the values:  \rangeAge~Gyr, stellar
metallicity,   \Z:  \rangeMetallicity,  the   e-folding  time   of  an
exponentially decreasing star formation history, \efold: \rangeTau~Gyr
\  and the  color  excess, \ebv:  \rangeEbv~mag.   The resultant  (4D)
initial  model grid  is  consist of  \ninitialmodel \  dust-attenuated
integrated stellar  spectra. The initial grid is  fixed throughout the
analysis.

\begin{figure}
\includegraphics[width=88mm]{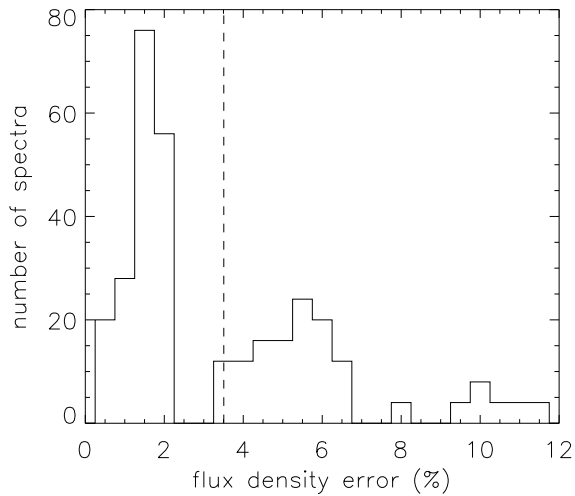}
\caption{The  distribution of  the error  in the  flux density  of the
interpolated  spectra  over a  4D  grid,  with  respect to  the  true
theoretical        spectra       from        the       dust-attenuated
\citet{2003MNRAS.344.1000B} model.  The vertical dashed line indicates
the average error, \avgferr.}
\label{fig:fluxerror}
\end{figure}

   An     example     interpolated     spectrum    is     shown     in
\figname\ref{fig:demointerpspec}.  Its  \age, \Z, \efold \  and \ebv \
are respectively \targetparm.  The difference spectrum, interpolated -
true theoretical spectra, is also shown. The true theoretical spectrum
is    referred   to    that   being    output   directly    from   the
\citet{2003MNRAS.344.1000B} program, and post dust-attenuated with the
\citet{2000ApJ...533..682C} model. The flux density error is driven by
both the parameter resolution in  the initial grid and the validity of
the linearity assumption within a  single bin of the initial grid.  In
the interpolated spectrum, the error is about 2\%, assigned equally to
each and every  wavelength bin to give a reduced  \chisq \ between the
two spectra  close to  unity (equals \examplerchisq  \ in  this case).
Extending to the  full parameter space, we generate  \nspecC \ spectra
at  \age \  = \rangeAgeC~Gyr,  \Z \  = \rangeMetallicityC\footnote{The
stellar metallicity values chosen here are not different from those in
our  initial grid,  because other  metallicity values  are  not output
directly  from the  \citet{2003MNRAS.344.1000B} program.   So  in this
example  we are  in  fact considering  a  3D interpolation  in a  4D
hypercube.}, \efold \  = \rangeTauC~Gyr \ and \ebv  \ = \rangeEbvC~mag
using the  above-mentioned initial grid,  which are compared  with the
true  theoretical   spectra.   \figname\ref{fig:fluxerror}  shows  the
distribution of  the flux density  error in the  interpolated spectra,
showing an average  of \avgferr, a minimum of  \minferr, and a maximum
of \maxferr;  with a  resultant average \chisq  \ of  \avgrchisq.  The
amplitude  of the  flux density  error is  comparable with,  e.g., the
spectrophotometric uncertainty from the  Sloan Digital Sky Survey Data
Release    6,   7\%    in   the    observed   frame    wavelength   of
3800~\AA~\citep{2008ApJS..175..297A}.   The   threshold  in  the  flux
density  error, at which  the interpolated  spectrum is  still useful,
would depend on individual applications.   We plan to extend this type
of comparison to  other models.  One model to  consider, e.g., is that
by  \citet{2009MNRAS.396..462K},   who  provide  an   interactive  web
interface  for downloading theoretical  spectra of  integrated stellar
systems which are defined by multiple parameters.

We next construct  a mock galaxy sample for  the parameter estimation,
in which the parameters underlying  each spectrum are known.  Based on
the initial grid, a random sample of \nsynthetic \ theoretical spectra
are generated through our approach.  Random Gaussian noise is added to
the flux  density in  each wavelength bin  of each spectrum,  fixed at
signal-to-noise (\sn) of  \snsynthetic.  Bayesian parameter estimation
is then performed on the  mock sample, using the approach described by
\citet{2007MNRAS.381L..74K}, where  all of the 4  model parameters are
estimated simultaneously.   The parameter estimations  are carried out
using several  different samples of theoretical  spectra, defined from
high     to    low     parameter    resolution,     as     shown    in
\tabname\ref{tab:testgrid}.   These spectra  are also  generated using
the model generation approach described in this work, and are based on
the  same initial  grid. As  such, we  are probing  the effect  on the
parameter  estimates due  to the  increasing targeted  grid resolution
only,  rather  than  that  due  to  model  imperfection  in  terms  of
describing  the real  observed spectra,  and  so on.   Except for  the
stellar metallicity,  for which we  follow \citet{2005MNRAS.362...41G}
and  adopt  logarithmic bins  (i.e.,  evenly  sized in  $\log_{10}Z$),
linear bins  (i.e., evenly sized bins)  are used for all  of the other
parameters (\age, \efold \ and \ebv) in the targeted grid.

\begin{table}
\caption{Tested parameter grids.}
\label{tab:testgrid}
\begin{tabular}{llr}
\hline
Targeted grid size & Reduction factor & Number of spectra \\
\hline
14 $\times$ 36 $\times$ 15 $\times$ 17 & 1   & 128520 \\ 
10 $\times$ 24 $\times$ 10 $\times$ 12 & 1.5 & 28800 \\ 
 7 $\times$ 18 $\times$  8 $\times$  9 & 2   & 9072 \\
 4 $\times$  9 $\times$  4 $\times$  5 & 4   & 720 \\
\hline
\end{tabular}

\medskip{There are 4 parameters  (\age, \Z, \efold, \ebv) constituting
  each of the  grids, in that order the grid size  is given. The range
  of  a  parameter  is  the  same  for all  of  the  grids,  given  in
  \S\ref{S:gridres}.    The  reduction   factor  is   the  approximate
  fractional  decrease of  grid  resolution relative  to  the case  of
  highest  resolution, and  is set  to be  uniform in  each parameter.
  Except in \Z  \ for which logarithmic bins  are adopted, linear bins
  are used in all of the other parameters.}
\end{table}

\begin{figure}
\begin{subfigmatrix}{1}
     \subfigure[The   age  of   the   oldest  stars.]{   
          \includegraphics[width=.24\textwidth,angle=0]{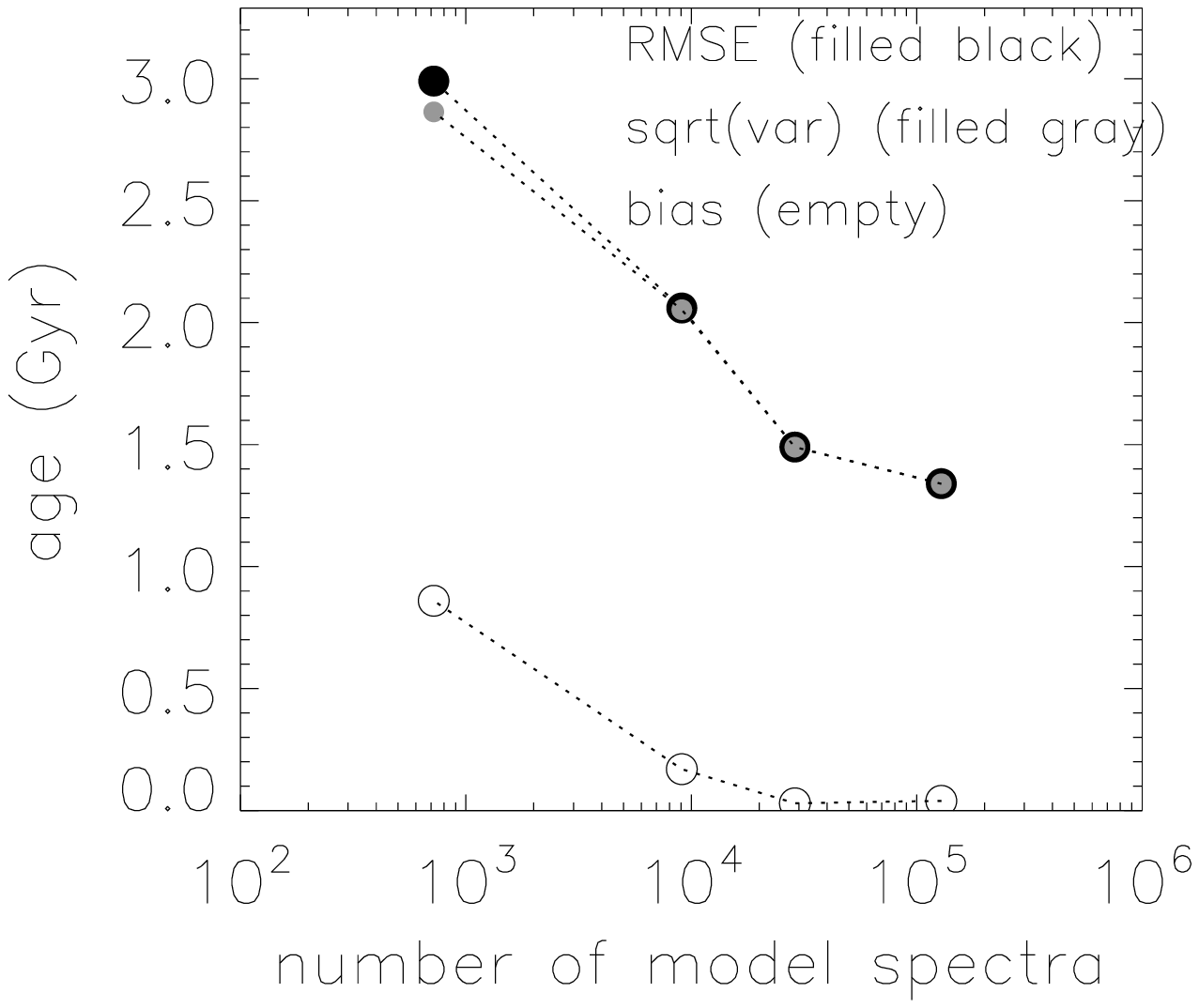}}
     \hspace{-0.3in}
     \subfigure[The    stellar    metallicity.]{    
       \includegraphics[width=.24\textwidth,angle=0]{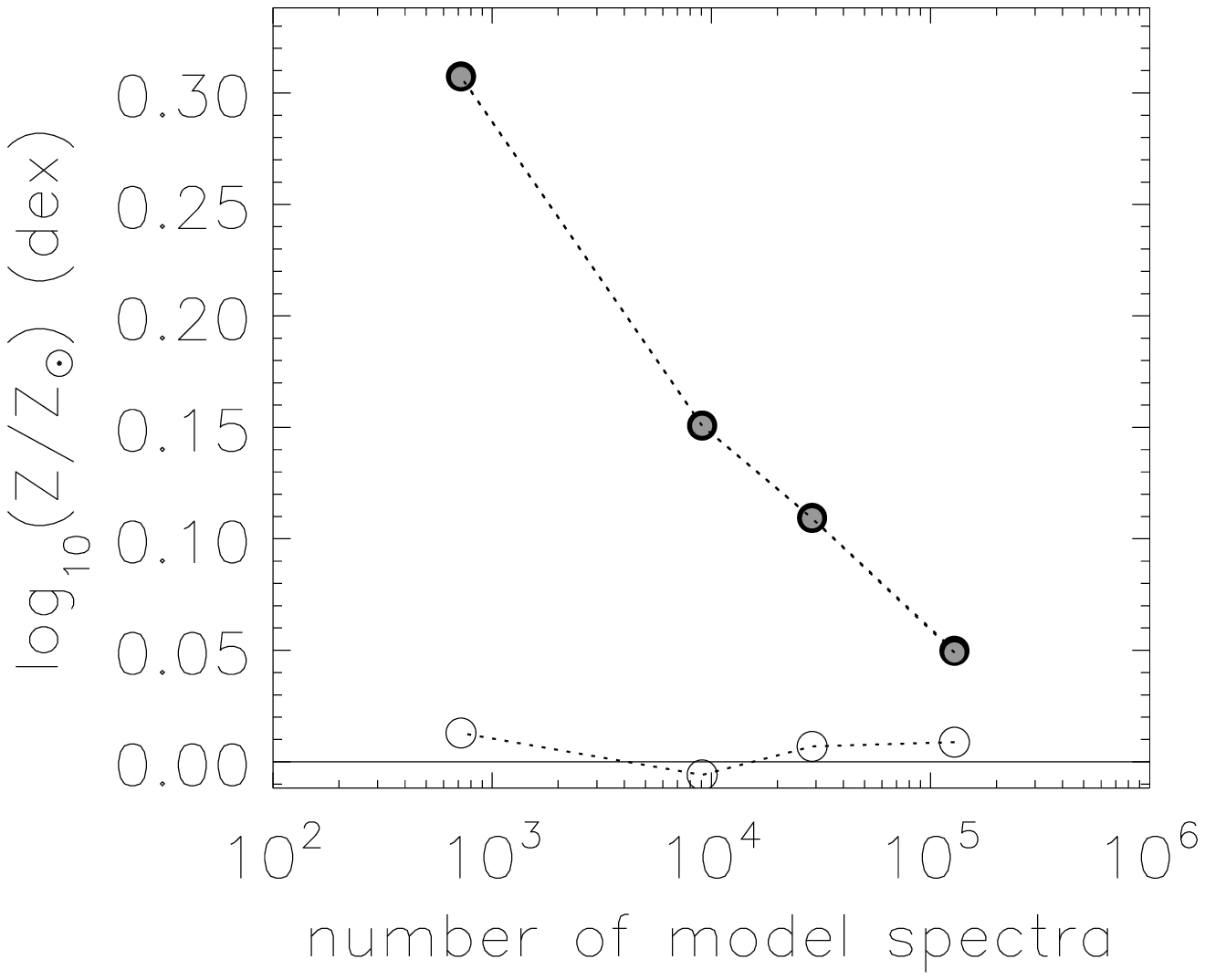}}
     \subfigure[The   e-folding   time   in the exponentially decreasing
     star   formation rate.]{
       \includegraphics[width=.24\textwidth,angle=0]{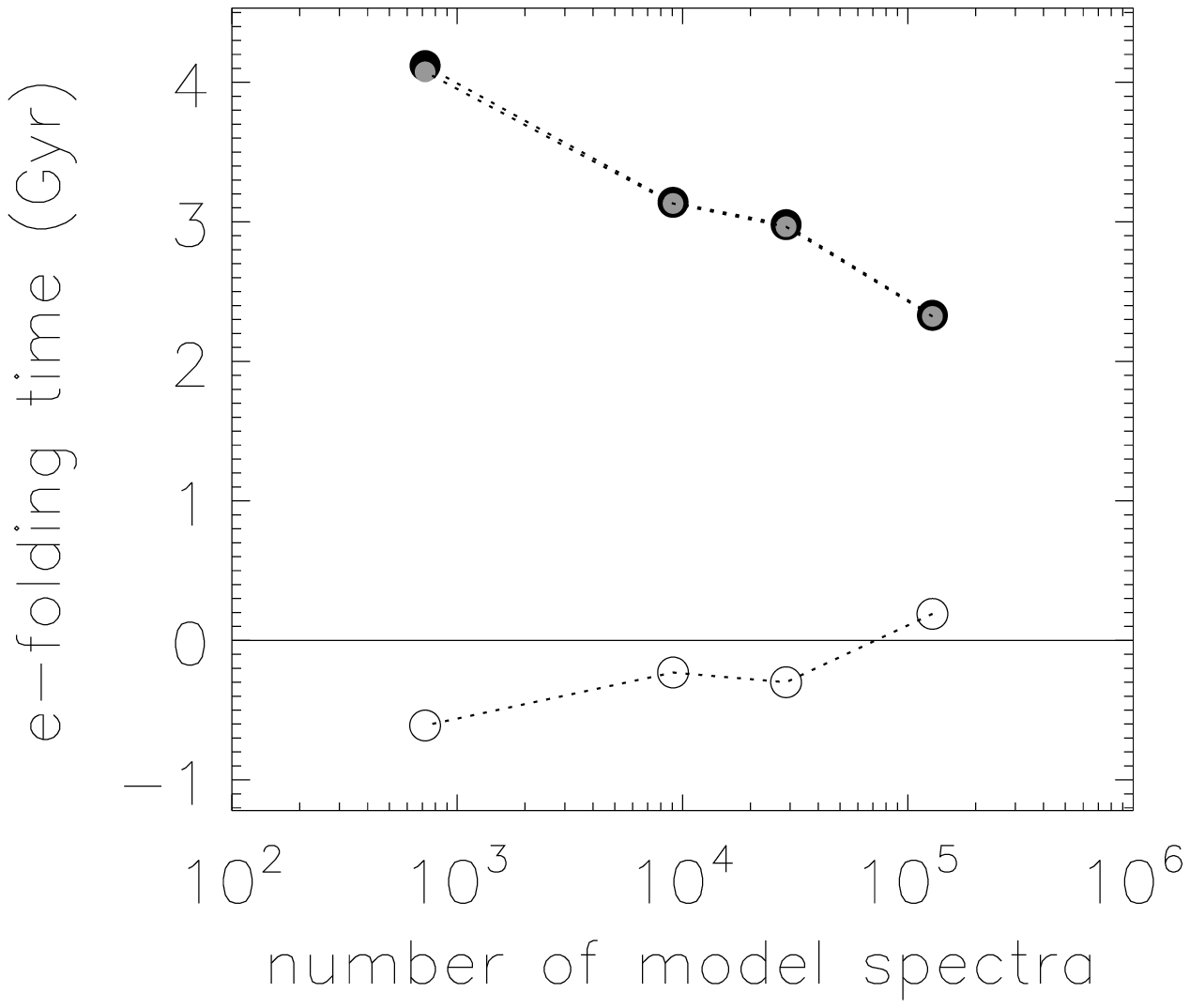}}
     \hspace{-0.3in}  \subfigure[The   color  excess.]{  
     \includegraphics[width=.24\textwidth,angle=0]{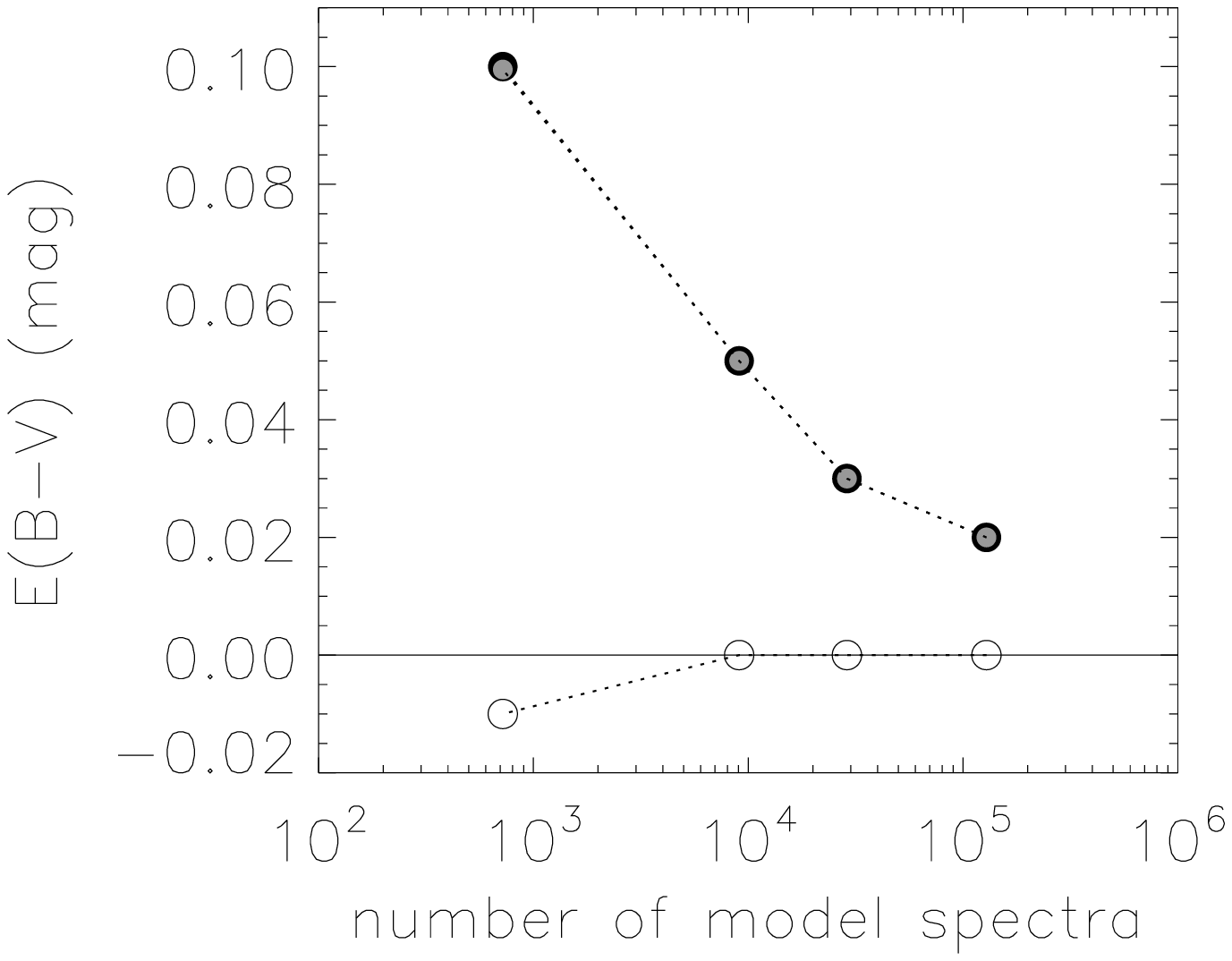}}
\end{subfigmatrix}
\caption{The dependence of the RMSE (filled black circles), the square
root  of  the variance  (filled  gray  circles)  and the  bias  (empty
circles)  of the  parameter  estimates on  the  number of  theoretical
spectra,  or  the resolution  of  the  targeted  parameter grid  (see
\tabname\ref{tab:testgrid}).  The solid horizontal lines mark the zero
value.  The dotted lines are for eye-guiding only. }
     \label{fig:rmsbias}
\end{figure}

The dependence of  the root mean square error  (RMSE), the square root
of  the variance  and the  bias  (the average  difference between  the
estimate and the true parameter  value) of the estimated parameter are
shown  in  \figname\ref{fig:rmsbias},  for  each  of  the  parameters.
Mathematically, the RMSE  is the square root of  the mean square error
(MSE).  The MSE is the sum of the variance and the bias squared, so it
can  be interpreted  as  the sum  of  the squares  of statistical  and
systematic errors \citep{1998Cowan}.  The  MSE is found to decrease as
the resolution of the targeted parameter grid increases, an indication
of  an improved determination  of the  posterior probability  for each
parameter of the  mock galaxies, in this case, due  to the increase in
the number of  parameter grid points being sampled.   We also see that
the  estimates are  fairly  unbiased when  the  number of  theoretical
spectra  \unbiasednspec   \  for  our  parameter   ranges,  where  the
corresponding grid resolution is listed in \tabname\ref{tab:testgrid}.
On the other  hand, the variance of the  estimates at our large-sample
limit   should   be  limited   by   the   data   \sn.   For   example,
\citet{2005MNRAS.362...41G} compared (their \figname~3) both the shape
and  the width  of the  posterior probability  in  stellar metallicity
between the  high- and  low-\sn \ spectra  from the Sloan  Digital Sky
Survey \citep{2000AJ....120.1579Y}, in that lower \sn \ spectrum gives
wider  distribution  in   the  posterior  probability  of  parameters.
Earlier \citet{2002AJ....123.1864G}  also came to  similar conclusion,
through   parameter  recovery  test   upon  synthetic   photometry  of
integrated stellar populations.

Another  interesting   aspect  of  \figname\ref{fig:rmsbias}   is  the
simultaneous improvement  in all of  the parameter estimates  upon the
increase  in  the targeted  parameter  grid  resolution.  This  result
manifests  the presence  of  degeneracies among  all  of the  involved
parameters  -- \age, \Z,  \efold \  and \ebv.   It is  well-known that
age-metallicity  degeneracy may  hinder the  studies  of astrophysical
objects by using only their broadband colors.  Even integrated spectra
are not immune from this  degeneracy -- an integrated stellar spectrum
with intermediate  to old age  (1.5~Gyr and up)  may look the  same by
tripling the  age or reducing the  stellar metallicity by  a factor of
two  \citep{1994ApJS...95..107W}.  To  determine  \age, one  therefore
needs to know  \Z, and vice versa, meaning both  parameters have to be
determined   simultaneously.   \figname\ref{fig:rmsbias}  demonstrates
that degeneracies exist among all of the considered parameters, in the
way that the improvements on these estimates go hand-in-hand.

\section{Summary} \label{S:summary}

In   applications   where   a   large  (sample   size   \largesample),
non-analytical model  is needed,  the hypercube representation  of the
parameter space, together with multi-linear interpolation for deriving
the theoretical value at an intermediate grid point, are shown here to
be feasible to improve the  flexibility in handling the models.  Using
stellar population  synthesis of galaxies  as a pilot study,  we found
that the  increase in targeted parameter grid  resolution improves the
parameter estimates in terms of the mean square error. This example is
given by incorporating  the Bayesian parameter estimation.  Naturally,
our  model generating approach  can be  combined with  other parameter
estimation techniques, such as the Markov Chain Monte Carlo.

Models of large sample size can be applied to studies in various areas
and disciplines.  In the context of spectral analyses, a large grid of
theoretical  spectra is  expected  to be  applicable  to mock  catalog
construction, studying the  relationship between spectral features and
model parameters,  spectral fitting, and  stellar population synthesis
of galaxies.   Our approach is particularly suitable  for studies that
are  based on multi-dimensional  parameter space,  and are  focused on
investigating  differences among  results  obtained through  different
models.

\section{Acknowledgments}

I thank  Rosemary~F.~G.~Wyse for discussions which lead  to this work,
and her  comments on the  manuscript.  I thank  Tam\'as~Budav\'ari and
Andrew~Ptak for discussions on parameter estimations, and Sandra~Faber
for discussions  on relating spectra to physical  parameters.  I thank
Niraj~Welikala  and Alex~S.~Szalay for  various discussions.   I thank
the referee for useful comments and suggestions. I acknowledge support
through grants from the W.~M.~Keck Foundation and the Gordon and Betty
Moore Foundation, to establish  a program of data-intensive science at
the Johns Hopkins University.

This research has made use  of data obtained from or software provided
by  the US  National Virtual  Observatory, which  is sponsored  by the
National Science Foundation.

{}

\end{document}